\documentstyle[12pt]{article}
\newcommand{\be}{\begin{equation}}
\newcommand{\ee}{\end{equation}}
\newcommand{\bea}{\begin{eqnarray}}
\newcommand{\ena}{\end{eqnarray}}
\newcommand{\beano}{\begin{eqnarray*}}
\newcommand{\enano}{\end{eqnarray*}}
\newcommand{\sect}[1]{\setcounter{equation}{0}\section{#1}}
\newcommand{\vs}[1]{\rule[- #1 mm]{0mm}{#1 mm}}
\newcommand{\hs}[1]{\hspace{#1 mm}}
\newcommand{\prt}{\partial}
\newcommand{\half}{{\frac{1}{2}}}
\newcommand{\shalf}{\mbox{$\frac{1}{2}$}}
\newcommand{{\cg}}{\mbox{$\cal{G}$}}
\newcommand{{\ch}}{\mbox{$\cal{H}$}}
\newcommand{{\co}}{\mbox{$\cal{O}$}}
\newcommand{{\cs}}{\mbox{$\cal{S}$}}
\newcommand{{\ci}}{\mbox{$\cal{I}$}}
\newcommand{{\cw}}{\mbox{$\cal{W}$}}
\newcommand{{\cws}}{\mbox{$\cal{W}_{\cal{S}}$}}
\newcommand{\cu}{\mbox{$\cal{U}$}}

\newcommand{\mb}[1]{\hs{5}\mbox{#1}\hs{5}}
\newcommand{\bal}{{\bar{\alpha}}}
\newcommand{\bbet}{{\bar{\beta}}}

\newcommand{\R}{\mbox{\hspace{.2mm}\rule{0.2mm}{2.8mm}\hspace{-1.5mm} R}}
\newtheorem{defi}{Definition}

\newtheorem{coro}{Corollary}
\newtheorem{prop}{Proposition}

\newcommand{\NP}[1]{Nucl.\ Phys.\ {\bf #1}}
\newcommand{\PL}[1]{Phys.\ Lett.\ {\bf #1}}

\newcommand{\CMP}[1]{Comm.\ Math.\ Phys.\ {\bf #1}}

\newcommand{\IJMP}[1]{Int.\ Journ.\ Mod.\ Phys. {\bf #1}}

\begin{document}
\renewcommand{\thefootnote}{\fnsymbol{footnote}}
\setcounter{page}{0}
\pagestyle{empty}

\vs{30}

\begin{center}
{\LARGE {\bf Finite \cw-algebras and intermediate statistics }}\\[2cm]

{\large F. Barbarin$^1$, E. Ragoucy$^1$, P. Sorba$^{1,2}$}\\[.5cm]

{\em {Laboratoire de Physique Th\'eorique}}
{\small E}N{\large S}{\Large L}{\large A}P{\small P}
\footnote{URA 14-36 du CNRS, associ\'ee \`a l'E.N.S. de Lyon,
et au L.A.P.P. (IN2P3-CNRS) d'Annecy-le-Vieux.\\
$^1$ Groupe d'Annecy, LAPP, Chemin de Bellevue, BP 110,
F - 74941 Annecy-le-Vieux Cedex, France\\
$^2$ Groupe de Lyon, ENS, 46 all\'ee d'Italie, F-69364 Lyon
Cedex 07, France}
\end{center}

\vs{15}

\centerline{\bf Abstract}

\vs{5}

New realizations of finite \cw-algebras are constructed by relaxing the usual
constraint conditions. Then finite \cw-algebras are recognized in the
Heisenberg quantization recently proposed by Leinaas and Myrheim, for a system
of two identical particles in $d$ dimensions. As the anyonic parameter is
directly associated to the \cw-algebra involved in the $d=1$ case, it is
natural to consider that the \cw-algebra framework is well adapted for a
possible
generalization of the anyon statistics.

\vfill

\begin{center}
{\it in memory of our friend and colleague Tanguy {\large\it Altherr}}
\end{center}

\vfill
\rightline{hep-th/9410114}
\rightline{{\small E}N{\large S}{\Large L}{\large A}P{\small P}-AL-489/94}
\rightline{September 1994}

\newpage
\pagestyle{plain}
\renewcommand{\thefootnote}{\arabic{footnote}}
\setcounter{footnote}{0}
\sect{Introduction\label{intro}}

\indent

Finite \cw-algebras have been introduced \cite{1,2} by studying symplectic
reductions of finite dimensional simple Lie algebras in complete
analogy with usual \cw-algebras constructed as reductions of affine
Lie algebras. It is reasonable to think, with the authors of \cite{2},
that a good understanding of the finite case can help to draw
informations on the infinite dimensional one. In the same time, one
must not be surprised if some of such finitely generated but non
linear algebras already appeared in theoretical physics. That is the
case of the finite \cw-algebra obtained from the reduction of the $sp(4,\R)$
model associated with the $s\ell(2,\R)$ embedding of Dynkin index 2
(i.e. built on the short root of $sp(4,\R)$ ). The corresponding
\cw-algebra is four dimensional, and can be seen as a "deformed"
version of $g\ell(2,\R)$ with a cubic term in the r.h.s. of
a commutation relation (see below). As remarked in \cite{3}, this
algebra has already been considered by P.W. Higgs as an algebra of
conserved quantities for a Coulombic central force problem in a space
of constant curvature \cite{4}.

\indent

We have recognized the same finite \cw-algebra in a completely
different framework, namely intermediate statistics. Indeed, in \cite{5},
J.M. Leinaas and J. Myrheim have considered the Heisenberg quantization of a
system of two identical particles in one and two dimensions. The
one-dimensional system can be formally related to a system of two
identical vortices in a thin, incompressible superfluid film, the two
spatial coordinates of the vortex center acting as canonically
conjugate quantities. Such a model has been proposed as anyon
candidate. The authors have remarked that one-dimensional Heisenberg
quantization is closely related to Schr\"{o}dinger quantization in two
dimensions, group theory playing an important r\^{o}le in the former
approach. Indeed, in the Heisenberg framework, intermediate statistics
\cite{Ler}
are clearly characterized by a continuous parameter, itself related to
the Casimir eigenvalue of the $sp(2,\R)$ observable algebra. The boson
and fermion cases correspond to limiting values of this parameter.

The situation is more complicated in the two-dimensional Heisenberg
quantization, from the algebraic point of view as well as from the
physical interpretation. The algebra of interest in now $sp(4,\R)$ and
more than one extra parameter are necessary to characterize the
different configurations. Although many
points still need to be clarified for a
generalization of the anyon notion, one remarks that in the two
cases considered above, the physical
interpretation is brought by a finite \cw-algebra. In the one
dimensional case, it is just the $s\ell(2,\R)$ Casimir operator, while
in the two-dimensional one, it is exactly the finite \cw-algebra
already described in \cite{3} and \cite{4}. Reconsidering the problem
in the \cw-algebra framework can help a lot for setting more rapidly
and systematically the algebraic formalism used in \cite{5}.

It is one of the purpose of this paper to present this approach,
which might be of some use to go further in the attempt of \cite{5} to
generalize intermediate statistics.

\indent

Moreover, the determination of the finite \cw-algebras in the
above examples has to be done without putting to constant the current
components relative to the constraints in the usual Hamiltonian
reduction. In other words, these components are left free,
although they still determine the transformations under which
the $W$ generators will be invariant. Only the case where such
components are commuting (or in other words the gauge group is
Abelian) will be considered here. Then, as will be shown on examples,
the $W$ generators depend explicitely on all the current components,
but their polynomiality property (e.g. in the highest weight gauge)
is lost. This property can be restored by enlarging the \cw-algebra
to a bigger algebra containing also the component formerly related to
the constraints (i.e. dual to the group transformations).

Such a construction of finite \cw-algebras by
keeping free the components associated to the constraints is the
second point that we would like to emphasize in this paper. Of course,
we hope that such a feature of finite \cw-algebras could also be extended
to the affine case, thus proving, following the previous assertions,
that finite \cw-algebras constitute a suitable laboratory for more
elaborated structures.

\indent

The paper is constituted by two sections. In the first one, we realize finite W
generators
as invariant quantities under Abelian subgroups $G_+$ of a simple group $G$.
Some of their properties are discussed, in particular the
correspondence, after quantization, between W and the commutant of
the Abelian subalgebra $\cg_-$. These technics
appear well adapted to study the representations of the Lie algebra
$sp(2d,\R)$ to which reduces, following \cite{5}, the Heisenberg quantization
for a system of two identical particles in $d$ dimensions. As in \cite{5}, the
$d=1$ and $d=2$ cases are considered in detail, with emphasis each time on the
r\^{o}le of the associated finite \cw-algebra, while general features for $d>2$
are rapidly mentioned.
Finally we suggest an alternative to the conclusion of \cite{5} in order to
recognize, through the \cw-algebra in the $d=2$ case, the anyonic parameter
well established for $d=1$.

\sect{New realizations of finite \cw-algebras\label{s2.0}}

\indent

We recall in the first paragraph some features of the construction of finite
\cw-algebras from Lie algebras. Then, we
present new realizations for these \cw-algebras. In opposition to the usual
Hamiltonian reduction, no constraints are necessary, so that these
realizations use all the generators of
the Lie algebra that lies behind the finite \cw-algebra. Finally, we quantize
the previous classical approach.

\subsection{The traditional Hamiltonian reduction \label{Wcont}}

\indent

As developed in \cite{1,2}, Hamiltonian reduction of Poisson
structures on simple Lie algebras can be performed to construct finite
\cw-algebras, in complete analogy with the affine case, the only
difference being the absence of the space variable.

\indent

We start with a simple, real, connected and maximally non compact
Lie group $G$, with Lie algebra \cg. Let $t^a$ be the basis of \cg,
and $J_a$ the dual basis in $\cg^*$:
\be
[t^a,t^b]= f^{ab}_c\ t^c \mb{} J_a(t^b)=\delta^b_a \label{CR(g)}
\ee
We introduce the metric on \cg\ in a representation $R$:
\be
\eta^{ab}=<t^a,t^b>= tr_R(t^at^b) \mb{and} \eta_{ab}\eta^{bc}=\delta_a^c
\label{metric}
\ee
One can define on $\cg^*$ a Poisson-Kirillov structure which mimicks
the commutators (\ref{CR(g)}):
\be
\{J_a,J_b\}= f_{ab}^c\ J_c \mb{with} f_{ab}^c=\eta_{ad}\eta_{be}\eta^{cg}
f^{de}_g \label{PB(g*)}
\ee

Now, as usual \cite{Bais,ORaf}, we introduce
a gradation on $\cg=\ \cg_-\oplus \cg_0\oplus \cg_+$ relative to an
$s\ell(2,\R)$ embedding w.r.t. which the reduction will be performed.
We will call $G_{0,+,-}$ the subgroups associated to $\cg_{0,+,-}$.
In order not to overload the presentation, let us limit our discussion to the
integral grading case. The half-integral gradations need some extra
precautions: we will add a few remarks on it in the next paragraph.
As a notation we use the indices $\alpha,\beta,\dots$ for
negative grades, $i,j,k,\dots$ for the zero grades, and
$\bal,\bbet,\dots$ for the positive ones. The corresponding sets will
be denoted by $I_{\pm,0}$. We keep $a, b, c, \dots$ for the
general indexation, and  $t^{0,+,-}$ are the
generators of the $sl(2,\R)$ under consideration.

Now, we introduce on $\cg^*$ first class constraints relative to the
$\cg_-^*$ part:
\be
J_\alpha-\chi_\alpha=0,\ \ \forall\alpha\in I_-
\ee
where $\chi_\alpha$ is a constant, which is zero
except when $J_\alpha=J_-$, the negative $s\ell(2,\R)$ root generator. We
take $\chi_-=1$ for simplicity.
The (first class) constraints weakly commute among themselves:
\be
\{J_\alpha-\chi_\alpha,\ J_\beta-\chi_\beta\}_{Const}=0 \label{chi}
\ee
where the index $Const$ indicates that one has to apply the constraints
after computation of the Poisson Bracket (PB). Following Dirac prescription,
these first
class constraints generate a gauge invariance on the $J_a$'s:
\be
J_a\ \rightarrow\ exp( c^\alpha \{ J_\alpha,.\}_{Const})(J_a)
=J_a+ c^\alpha \{J_\alpha,J_a\}_{Const}+
\half c^\alpha c^\beta \{J_\beta,\ \{J_\alpha, J_a\}\}_{Const}+ \dots
\label{dual}
\ee
where the $c^\alpha$ are the gauge transformation parameters.

It is useful to keep in mind that these gauge transformations can be
performed either through commutation relation on \cg, or via PB in
$\cg^*$. Indeed, let us introduce the usual constrained matrix
\be
J= t^{-} + J_it^i + J_{\bar{\alpha}}t^{\bar{\alpha}} \label{Jcont}
\ee
and look at the gauge transformation (\ref{dual}):
\be
J\ \rightarrow\ J^g=exp( c^\alpha \{ J_\alpha,.\}_{Const})(J)
\ee
Developing $J^g$ with the help of the gradation and the use of the
constraints, we can, thanks to the relations
(\ref{CR(g)}-\ref{PB(g*)}), rewrite $J^g$ as
\be
J^g=exp([c_\bal\ t^\bal,.])(J)=g^{-1}Jg \label{group}
\ee
where we have introduced the group element $g=exp(c_\bal\ t^\bal)$ and
the parameters $c_\bal=\eta_{\bal\alpha}c^\alpha$. Thus, the gauge
transformations can be seen as conjugations on \cg\ by elements
of the subgroup $G_+$.
Note that at the affine
level, the above discussion is still valid but, because of the central
term in the PB (\ref{PB(g*)}), one finds the (usual) coadjoint transformation
instead of conjugation. Here, the differential term being identically zero,
the conjugation and the coadjoint transformation are identical.

\indent

Finally, one fixes the gauge (\ref{group}) by demanding the transformed
current $J^g$ to be of the form (highest weight gauge)
\be
J^g= t^- +W_s\ \ell^s \mb{with} [t^+,\ell^s]=0\ ;\ [t^0,\ell^s]=s\ \ell^s
\ee
The $W_s$'s are in the enveloping algebra of $\cg^*$, and they are
gauge invariant. They form a basis of the finite \cw-algebra. Of
course, by construction, the $W$ generators, which are the gauge
invariant polynomials in the $J_i$'s and $J_\bal$'s will have
{\it weakly} vanishing PB with the constraints.

\subsection{Relaxing the constraints\label{s2.2}}

\indent

Instead of using the constrained current (\ref{Jcont}), one can
perform the coadjoint transformations on the complete matrix
$J_{tot}=J_a\ t^a$:
\be
J^g_{tot}=\ g^{-1}J_{tot}\, g \mb{with} g\in G_+\label{Jtotg}
\ee
Then, developing $J_{tot}^g$ using the same
rules as in the previous section, we get
\be
J_{tot}^g= exp(c^\alpha \{ J_\alpha,.\})(J_{tot}) \label{Jtotg*}
\ee
where now the PB are computed {\it without} any use of what were
the constraints. Thus, if one finds (as we will show below)
quantities which are invariant
under the coadjoint transformations, these objects will have
{\it strongly} vanishing PB with the generators $J_\alpha$.

Note that,
although the transformations we are looking at have the same form as
the gauge transformations of the previous section, they are
{\it not} gauge transformations, since they are not associated
to constraints. In other words, one cannot construct a
action associated to the transformations (\ref{Jtotg}), while for the
previous section, the action
\be
S(g,A_+,A_-)= tr\int\ dx_0\ \frac{dg}{dx_0}\frac{d\ g}{dx_0}^{-1} +
tr\int\ dx_0\ A_+(g^{-1}\frac{dg}{dx_0}-t^-) +(\frac{dg}{dx_0}g^{-1}-t^+)A_-
+g^{-1}A_+gA_-
\ee
is really invariant under the gauge transformations
\be
g(x_0) \rightarrow g_+(x_0)\ g(x_0)\ g_-(x_0)
\ee
Thus, the construction we present is strickly algebraic, but we will
see in the next section
that the technics can be applied to physical problems.

\indent

We will limit our study to the case where the constraints (\ref{chi})
have vanishing PB, that is to say when the conjugation group $G_+$ is
Abelian.

Note that, if $G_+$ is Abelian, the \cg-gradation reduces to
\be
\cg=\ \cg_{-1}\oplus\cg_0\oplus\cg_{+1} \label{-101}
\ee
Indeed, (\ref{-101}) insures that $\cg_{-1}$ is Abelian. Conversely,
as we are considering gradation w.r.t. an $s\ell(2,\R)$ Cartan generator,
the simple roots have grades $0\leq h\leq1$ only. Thus, a grade -2
generator (if it exists) is not a simple root generator, and must
be obtained as the commutation of two simple root generators of grade
-1, so that $\cg_{-1}$ would not be Abelian. An analogous reasoning
for half gradations leads to
\be
\cg=\cg_{-1}\oplus\cg_{-\half} \oplus\cg_0\oplus\cg_{\half}\oplus\cg_{+1}
\label{-.50.5}
\ee
Let us add that in that case, using the halving technics \cite{ORaf,Bais}, one
can split $\cg_{-\half}$ in such a way that the second class
constraints become first class. In most of the cases, an half-integral
gradation can be redefined into an integral one by using an
$s\ell(2,\R)$-commuting $U(1)$ factor \cite{U1}.

Using \cite{FRS}, it is easy to see that the gradations of types
(\ref{-101}-\ref{-.50.5}) correspond to a reduction w.r.t.:

$(i)$ the diagonal $s\ell(2,\R)$ in $p$
commuting $s\ell(2,\R)_1$ (the subscript is the Dynkin
index, $1\leq p\leq [\frac{N}{2}]$) for $\cg=s\ell(N)$;

$(ii)$ $s\ell(2,\R)_1$ or the diagonal $s\ell(2,\R)$ in $p$ commuting
$s\ell(2,\R)_2$ ($1\leq
p\leq d$) for $\cg=sp(2d)$;

$(iii)$ the diagonal $s\ell(2,\R)$ in $p\
s\ell(2,\R)_1$ ($1\leq p\leq [\frac{N}{2}]$) for $\cg=so(N)$.

\indent

Now, coming back to the transformations (\ref{Jtotg}), it is easy to
see that, because $G_+$ is Abelian, the negative grade generators in
$J_{tot}^g$ are not affected by $g$:
\be
J^g_{tot}= J_\alpha\ t^\alpha+\ K_{\geq 0}
\ee
with $K_{\geq0}$ decomposing on the positive grades $\cg_0\oplus
\cg_+$ only:
$K_{\geq0}= K_i\ t^i +K_\bal\ t^\bal$.
Then, it is natural to define
\be
\tilde{t}^-= J_\alpha\ t^\alpha
\ee
This generator being nilpotent, it can be included in an $s\ell(2,\R)$
subalgebra\footnote{In our construction, we discard the case where
$\tilde{t}_-=0$.} $\{\tilde{t}_\pm, \tilde{t}_0\}$. Once the other generators
$\tilde{t}_+, \tilde{t}_0$  have been identified, we can
take as symmetry fixing a highest weight form:
\be
J^g_{tot}= J_\alpha\ t^\alpha+ \tilde{W}_s\ \tilde{\ell}^s \mb{with}
[\tilde{t}^+, \tilde{\ell}^s]=0 \label{eq.hw}
\ee
where the index $s\in I_{hw}$ labels the highest weight generators.
Then, we have the property:
\begin{prop}\label{prop1}
The set $\cs=\{\tilde{W}_s,\ J_\alpha;\ s\in I_{hw},\alpha\in
I_-\}$ defines an enlarged
finite \cw-algebra called $\cws$, closing
(w.r.t. PB) on rational functions of the form
$P(\tilde{W}_s,J_\alpha)/Q(J_\alpha)$, and
realized by quotients of polynomials in the
$J_a$'s over polynomials
only in the $J_\alpha$'s.

Moreover, the subset $\cg^*_-=\{ J_\alpha;\ \alpha\in I_-\}$
is in the center of the $\cws$ algebra.
\end{prop}

{\it Proof:}

It is clear, using the same arguments as for the usual
highest weight gauge \cite{ORaf},
that the highest weight form determines uniquely the
symmetry generator $g$.
Thus, $J^g_{tot}$ will provide a complete set
of symmetry invariant generators. One has to
be careful that these generators will not only be
the $\tilde{W}_s$ generators but also the $J_\alpha$'s. Moreover, from
the form of $\tilde{t}^-$ and the polynomiality of the usual highest
weight gauge, one deduces the $\tilde{W}_s$
generators are quotients of polynomials in the $J_a$'s over polynomials in
the $J_\alpha$'s only.
Note that since the $\tilde{W}_s$ are fractions (and no more
polynomials), the set \cs\ generates a complete set of invariant
generators through rational functions $P(J_a)/Q(J_\alpha)$ instead of
polynomials. However, we remark that any invariant polynomial
$P(J_a)$ can be rewritten as a polynomial $Q(\tilde{W}_s,J_\alpha)$. Indeed,
starting from an invariant polynomial $P(J_a)$, let us perform the
symmetry transformation $g$ that leads to the highest weight form.
Then, from the invariance of $P$ and the highest weight form, we get:
\be
P(J_a)=P(J_a)^g=P(J_a^g)=P(\tilde{W_s},J_\alpha,0)\equiv
Q(\tilde{W}_s,J_\alpha)
\ee
We note that the PB of two $\tilde{W}_s$ generators reads
\be
\{\tilde{W}_s', \tilde{W}_s''\}=
\frac{P(\tilde{W}_s,J_\alpha)}{Q(J_\alpha)} \label{toto}
\ee
We conjecture that we can get rid of the denominator $Q(J_\alpha)$ and
get a "reduced" polynomial $P_r(\tilde{W}_s,J_\alpha)$. If this
conjecture is true, then $\cws$ will close polynomially in the
$\tilde{W}_s$'s and $J_\alpha$'s.
In the section \ref{sN2d2}, we show
explicitly an example where the conjecture is verified.

Finally, because of the invariance, any element of \cs\ will commute
with the $J_\alpha$'s, so that $\cg^*_-$ is in the center of $\cws$.

\indent

It should be obvious that the original \cw-algebra (obtained by
Hamiltonian reduction) can be deduced
from the enlarged \cw-algebra just by applying the constraints that
change $\tilde{t}_+$ into $t_+$.
Moreover, the new generators being in the center, one can
consider the quotient by $\cg^*_-$. As we are looking at \cw\
algebras, the definition of the quotient has to be slightly modified:
\begin{defi}\label{defi1}
Let $\cws$ be a \cw-algebra and $\cs_0$ a subset of the center of
$\cws$, one defines the quotient $\cws/\cs_0$ through the
equivalence relation
\be
W\equiv W'\ \ \Leftrightarrow \ \ W=F.W'+G
\ee
where $F,G$ are smooth functions of the elements of
$\cs_0$ such that $F\neq0$.
\end{defi}

Starting from this definition, one can
consider the PB $\{[W], [W']\}'\dot{=}[\{W,W'\}]$, where $[W]$ is an
element of $\cws/\cg^*_-$. In this PB one can
set formally the $J_\alpha$'s to the constraints $\chi_\alpha$ of
section \ref{Wcont}, since they are no more present in the quotient. But
on the hyperplan $J_\alpha=\chi_\alpha$, the $\cws$ is the \cw\
algebra of the previous section. Thus, we have
\begin{prop}\label{prop2}
The quotient $\cws/\cg^*_-$ is the finite \cw-algebra associated to the
gauge group $G_+$.
\end{prop}

{}From proposition \ref{prop2}, we deduce
\begin{coro}\label{lemma}
Starting from the enlarged \cw-algebra associated to the symmetry
group $G_+$, there is a consistent way of "renormalizing" the
generators $\tilde{W}_s$ by smooth functions of the $J_\alpha$'s in
such a way that the renormalized generators $\tilde{W}_s$ form the
\cw-algebra associated to the group $G_+$.
\end{coro}

\subsection{Quantization\label{s2.3}}

\indent

The previous realizations use generators in the dual
space $\cg^*$. In the case of (usual) Hamiltonian reduction, it means
that we are working at the classical level. However, it is often
useful (as we will see) to have a realization of the \cw-algebra
on \cg\ itself, thereby quantizing (in the Hamiltonian framework case)
the system. For finite \cw-algebras, the work is easy since
there always exists an (algebra) isomorphism between \cg\ and its dual. After
reminding this isomorphism, we will extend it to the enveloping
algebra $\cu^*$ of $\cg^*$, and then apply it to the realization of
finite \cw-algebras.

\indent

Let $i$ be the isomorphism between $\cg^*$ and \cg:
\[
i(J_a)=t_a=\eta_{ab}t^b \mb{with} t_a(t^b)=<t_a,t^b>
\]
where $\eta_{ab}$ is the inverse matrix of the metric $\eta^{ab}$.
This isomorphism is a Lie algebra isomorphism that sends the PB into the
commutator:
\[
{[i(J_a),i(J_b)]}= \eta_{ad}\eta_{be}\ [t^d,t^e]= {f_{ab}}^c\ i(J_c)=
i\left(\left\{J_a,J_b\right\}\right)
\]
It allows to relate the coadjoint and the adjoint actions.
Indeed, the coadjoint action $Ad^*$ is defined by
\[
{[Ad^*(g)(J_a)]}(X)= J_a(Ad(g^{-1})(X)) \mb{} \forall X\in\cg \mb{with}
Ad(g)(X)= gXg^{-1}\, \ g\in G
\]
Using $i$ and the group-invariance of $<.,.>$ one gets
\[
<t_a,g^{-1}Xg>=<gt_ag^{-1},X>= <Ad(g)(t_a),X>
\]
so that
\[
i\ o\ Ad^* = Ad\ o\ i
\]

Now, we extend $i$ to the enveloping algebra
$\cu(\cg^*)\equiv\cu^*$ with the rule
\be
i(J_{a1}\cdot J_{a2}\cdots J_{an})= S(t_{a1},t_{a2},\dots,t_{an})
\mb{}\forall n
\ee
where $S(.,.,...,.)$ stands for the {\it symmetrized} product of the
generators $t_a$.
$S$ is normalized by $S(X,X,..,X)=X^n$.

Note that $i$ is a vectorspace isomorphism
between $\cu^*$ and the set $\cu_{Sym}$
of totally symmetric polynomials in the $t_a$'s.
This set is a algebra (w.r.t. the commutator),
but $i$ is no more an algebra isomophism. However,
$i$ satisfies the fundamental property
\be
i\left( \{J_a, P\} \right)= [ t_a, i(P)] \mb{} \forall\ J_a\in\cg^*,\forall\
P\in\cu^*
\ee
Then, this implies that the symmetry transformations are "preserved"
by $i$.

Now, starting with a {\it polynomial} $W\in\cws$, one can construct the
operator
$\overline{W}=i(W)\in \cu_{Sym}$. Since $W$ is invariant under the group
transformations (\ref{Jtotg*}), we deduce that $\overline{W}$
will be invariant under the transformations
\be
\overline{W} \rightarrow exp(c^\alpha[t_\alpha,.])(\overline{W})
\label{bWg}
\ee
Of course, the above construction can be done in $\cu_{Sym}$
{\it a priori} only for polynomials. However, as $\cg_-^*$ is in the
center of $\cws$, if one considers the subset $i(\cws)$ in $\cu_{sym}$,
quotients by polynomials $P(J_\alpha)$ will be allowed. Thus, $i(W)$
will be well-defined in $i(\cws)$ for any element in $\cws$.

{\em In other words, if $W$
has vanishing PB with the $J_\alpha$'s, then $\overline{W}$ will
commutes with the image of these generators, that is the
$t_\alpha$'s.

The set of $\overline{W}$ fields being a complete set of invariant
under the transformations (\ref{bWg}), it is clear that it will form a
quantized enlarged \cw-algebra.}

 Let us remark that although
$i$ is an algebra isomorphism between $\cg^*$ and \cg, it is not an algebra
isomorphism between $\cu^*$ and $\cu_{Sym}$. Thus, one
has to be careful that the
commutation relations between $\overline{W}$ generators may be
different from the PB between $W$ fields: an example of the difference
between the classical and quantized \cw-algebras can be found in
\cite{thTjin}.

\sect{Intermediate statistics and finite \cw-algebras\label{s3.0}}

\indent

As an application of the results of section \ref{s2.0},
we consider the Heisenberg
quantization for systems of two identical particles as developed in
\cite{5}. Although the one dimensional case is algebraically simple, it
deserves to be presented, first as a pedagogical example and secondly because
it insures the connection with the usual anyon statistics. Then we turn to the
$d=2$ case emphasizing the r\^{o}le of a finite \cw-algebra. A short paragraph
is
also devoted to the generalization to higher dimensions. We conclude by an
analysis which leads to recognize, in the $d=2$ more general framework, the
anyonic statistics parameter. In what follows, we naturally use, as often as
possible, the notations of \cite{5}.

\indent

\subsection{Two particles in one dimension}

\indent

The relative coordinate and momentum of the two particle system are denoted by:
\be
x=x_{(1)} - x_{(2)} \ \ \ \ \ \ p=\frac{1}{2} \left( p_{(1)} -p_{(2)} \right)
\label{eq:1}
\ee
and satisfy the C.R.:
\be
\{x,p\}=1 \ \ \ \ \ [x,p]=i
\label{eq:2}
\ee
in either the classical or the quantum case.

The chosen observables:
\be
E_+=\frac{1}{2}\ p^2 \ \ \ E_- =\frac{1}{2} \ x^2 \ \ \ H=\frac{1}{4i} (xp+px)
\label{eq:3}
\ee
close, in the quantum case, under the C.R.'s of the $\cg =sp(2,\R) \simeq
s\ell(2,\R)$-like algebra:
\be
[E_+,E_-] = 2H \ \ \ \ \ [H,E_{\pm} ] = \pm E_{\pm}
\label{eq:4}
\ee
The realization (\ref{eq:3}),
where as usual $x^2$ acts as a multiplicative factor and
$p=-i  \prt_x$, is associated to the eigenvalue
\be
\gamma_0 = - \frac{3}{16}
\label{eq:5}
\ee
of the Casimir operator:
\be
\Gamma= H^2 + \frac{1}{2} (E_+ E_- + E_- E_+)
\label{eq:6}
\ee

Following \cite{5}, we wish to use a "coordinate" representation: then one
natural observable to diagonalize is $x^2$.

It is certainly the most well known property that in $s\ell(2,\R)$, one can
diagonalize simultaneously one generator and the  Casimir invariant, and also
that these two generators constitute a complete set of commuting observables.
However, let us remark that this property can be reobtained by considering the
$2 \times 2$ matrix representation of $s\ell(2,\R)$:
\be
J=
\left(
\begin{array}{cc}
J_0 & J_+ \\
J_- & -J_0
\end{array}
\right)
\label{eq:6b}
\ee
and determining the finite \cw-algebra generated by $\cg^*$ quantities
invariant under the subgroup $G_+$ generated by $E_+$. Indeed, with a group
transformation
\be
J^g= g^{-1} J g \mb{with}
g=\left(\begin{array}{cc} 1 & a \\ 0 & 1 \end{array}\right)
\ee
we find in the highest weight basis
\be
J^g= \left(\begin{array}{cc} 0 & (J_+J_-+J_0^2)/J_- \\ J_- & 0
\end{array}\right)
\ee
Then, using the isomorphism
$i: {\cg}^* \rightarrow {\cg}$ as explained in section \ref{s2.3},
one will recover
the $s\ell(2,\R)$ Casimir operator as generating (besides $E_-$ itself)
the commutant of the
$E_-$-generator. Of course, this approach reminds us the realization, in the
affine case, of the $W_2$ -or Virasoro- algebra via the Drinfeld-Sokolov
construction, usually obtained with the constraint: $J_-=1$.

Now, one can modify the $s\ell(2,\R)$ realization (\ref{eq:3})
into still a $x$ and $p$ one,
but valid for any eigenvalue $\gamma$ of $\Gamma$. Indeed, let us involve
simultaneously the coordinate observable $E_- =\frac{1}{2} x^2$ and the Casimir
operator $\Gamma$. For such a purpose, we keep unchanged $E_-$, as well as
the expression of $H$, which can be seen as a function of $x^2$ and its
conjugate
$\prt_{x^2}$:
\be
H=-(x^2 \prt_{x^2} + \frac{1}{4})
\label{eq:7}
\ee

Using (\ref{eq:5}) and (\ref{eq:6}), and considering the difference $\Gamma
- \Gamma_0$, we deduce the shift on $E_+$:
\be
E_+ \rightarrow E'_+ = E_+ + (\gamma + \frac{3}{16})\frac{1}{x^2}
\label{eq:8}
\ee
which does not affect the $s\ell(2,\R)$ C.R.'s.

As discussed by the authors of \cite{5}, the parameter:
\be
\lambda = \gamma + \frac{3}{16}
\label{eq:9}
\ee
can be directly related to the anyonic continuous parameter, with end point
$\lambda =0$, or $\gamma = -\frac{3}{16}$, corresponding to the boson and
fermion cases.

\indent

\subsection{Two particles in two dimensions \label{sN2d2}}

\indent

The algebra under consideration is generated by the quadratic homogeneous
polynomials in the relative coordinates $x_j$ and $p_j\ (j=1,2)$. One can
recognize a realization of the $sp(4,\R)$ algebra, the generators of which can
be conveniently separated into three subsets:

- the three (commuting) coordinate operators:
\be
u = (x_1)^2 + (x_2)^2 \ \ \ \ \  v = (x_1)^2 -(x_2)^2 \ \ \ \ \  w= 2x_1x_2
\label{eq:10}
\ee

- the three (commuting) second order differential operators:
\be
U = (p_1)^2 + (p_2)^2 \ \ \ \ \ V=(p_1)^2 - (p_2)^2 \ \ \ \ \ W=2p_1p_2
\label{eq:11}
\ee

- the four first order differential operators, generating an $s\ell(2,\R)
\oplus
{g}\ell(1)$ algebra:
\be
\begin{array}{ll}
C_s = \frac{1}{4} \sum_{i=1}^2 (x_i p_i + p_i x_i) & C_d = \frac{1}{4} (x_1
p_1 + p_1 x_1 - x_2 p_2 - p_2 x_2) \\
L=x_1p_2 - x_2p_1 & M= x_1 p_2 + x_2 p_1
\end{array}
\label{eq:12}
\ee
$C_s$ being the Abelian factor.

\indent

Choosing the $4 \times 4$ matrix representation of the ${\cg} = sp(4,\R)$
algebra
\be
M=m^{ij} \ \ E_{ij} \ \ \ \ \ \ \ i,j=1,2,3,4
\label{eq:13}
\ee
with $m^{ij}$ real numbers satisfying
\be
m^{ij} = (-1)^{i+j+1} \ \ m^{5-j, 5-i}
\label{eq:14}
\ee
the most general element of $\cg$ writes:
\be
J= J_a t^a =
\left[
\begin{array}{cccc}
-2J_{C_s}+J_M & J_L-2J_{C_d} & -J_V & J_U - J_W \\
-J_L -2J_{C_d} & -2J_{C_s}-J_M & -J_U-J_W & J_V \\
J_v & J_u+J_w & 2J_{C_s} + J_M & J_L-2J_{C_d} \\
-J_u+J_w & -J_v & -J_L -2 J_{C_d} & 2J_{C_s} -J_M
\end{array}
\right]
\label{eq:15}
\ee

As in the $d=1$ case, the $(x_j,p_j)$ realization as given in
(\ref{eq:10}-\ref{eq:12}) is a particular one, associated with special
eigenvalues of the $sp(4,\R)$ fundamental invariants $\Delta^{(2)}$ and
$\Delta^{(4)}$. But one can proceed as before in order to get a more general
representation. Wishing to keep as observables the three commuting coordinate
generators $u,v,w$ (coordinate representation), one may think of determining in
the ${\cg} = sp(4,\R)$ enveloping algebra the commutant of these three
generators. A basis for this subalgebra can be obtained, in a systematic way,
via the technics of section \ref{s2.0}.
More precisely, such a commutant can be seen as
a finite enlarged \cw-algebra.

First, one remarks that the ${\cg}$ basis decomposition given in
(\ref{eq:10}-\ref{eq:12}) naturally defines a ${\cg}$-grading:
\be
{\cg} = {\cg}_{-1} \oplus {\cg}_0 \oplus {\cg}_1
\label{eq:16}
\ee
with ${\cg}_{-,+,0}$ generated by $\{u,v,w\}, \{ U,V,W \}$ and  $\{ C_s,
C_d, L
, M \}$ respectively. This grading is itself related to the $s\ell(2,\R)$
embedding
obtained by taking the diagonal part of the
$s\ell(2,\R) \oplus s\ell(2,\R)$ subalgebra of
$sp(4)$ \cite{3,FRS}. Note that the $s\ell(2,\R)_2$
negative root generator $J_-$ can be seen
on the matrix (\ref{eq:15}) by selecting the $\cg_-$ part, that is by replacing
by zero all the entries except the ones where stand $J_u,J_v,$ and $J_w$.
To compare with
the (usual) Hamiltonian reduction process, in this last framework the
constraints could be chosen as:
\be
J_u=J_v=0, \ \ J_w=1
\label{eq:17}
\ee

There exist $G_0$ elements connecting these two bases\footnote{We
recall that $\tilde{t}_-\neq0$}: let $T(J_u,J_v,J_w)$ one of
these elements:
\be
T\
\left(
\begin{array}{cccc}
0 & 0 & 0 &0\\
0 & 0 & 0 &0\\
0&1 &0&0\\
1&0 & 0&0
\end{array}
\right)
\ T^{-1}\ = \
\left(
\begin{array}{cccc}
0 & 0 &0 &0\\
0 & 0 &0 &0\\
J_v &  J_u+J_w & 0&0\\
-J_u+J_w &- J_v & 0 &0
\end{array}
\right)
\label{eq:18}
\ee

One knows that the decomposition of $\cg$ w.r.t.
$s\ell(2,\R)_2$ representations is:
\be
{\cg} = D_0 \oplus 3D_1
\label{eq:19}
\ee

Using $G_+$ transformations, it is just a question of computation to
determine in the highest weight basis the four quantities (one for each
$s\ell(2,\R)_2$
representation in $\cg$) which generate (besides $u,v,w$ themselves)
the set of invariants under this
Borel subgroup. The highest weights can be determined as explicited in
(\ref{eq.hw}), or by acting with the $T$ conjugation (\ref{eq:18})
on the highest
weights relative to the $s\ell(2)_2$ basis determined by (\ref{eq:19}).
Practically, one has to find $g_+ \in G_+$ such that:
\be
g^{-1}_+ Jg_+ = J_u \ u + J_v\ v + J_w\ w + \sum_s
\widetilde{W}_s\ \widetilde{l}^s
\label{eq:20}
\ee
keeping the notations of (\ref{eq:10}-\ref{eq:19}).

At the classical level, the $\widetilde{W}_s$ quantities Poisson commute with
$J_u, J_v$ and $J_w$. Using the $i$-transformation from $\cg^*$ to $\cg$ and
the approach described in section \ref{s2.3},
one can choose as generators of the
commutant of the Abelian subalgebra $\{u,v,w\}$ (in addition to $u,v,w$
themselves):
\bea
\hat{S} &=& {-uL + vM - 2wC_d} \\
\hat{R} &=& \frac{-1}{24} \left\{-v^3W + 4v^2 C_sL+ 4v^2 M
C_d +v^2 wV - vw^2W+ \right.\nonumber \\
&&\mb{} +2vwM^2+vu^2W - 4uvMC_s-4uvLC_d-8vwC_d^2
+8uwC_d C_s - wu^2V+\nonumber \\
&&\mb{} \left. +w^3V -2uwLM+4w^2LC_s-4w^2 C_d M + Sym.\right\} \\
\hat{Q} &=& \frac{1}{120} \left[ \left\{ 4u^2wLC_d-u^3w
W+uw^3W+v^3uV-2v^3ML+ \right. \right. \nonumber \\
&&\mb{} + v^2 uwW + 4v^2wC_d L -8uvwC_dM+v^4U+w^4U
-4v^2wMC_s+2uv^2M^2+\nonumber\\
&& \mb{} +2uv^2L^2 +4w^3 C_d L +8uw^2 C_d^2 + 2w^2uL^2-8vw^2 C_d C_s
-2 v u^2 ML -2w^2vML+ \nonumber \\
&&\mb{} +2v^2 w^2U -u^2 v^2U - u^2w^2U -8v^3C_sC_d-4w^3 C_s M
-vu^3 V + v uw^2V+ \nonumber \\
&& \left.\left.\mb{} +8vu^2 C_s C_d + 4u^2w C_s M + Sym. \right\}
 -\frac{28}{3}u (u^2-v^2-w^2) \right] \\
\hat{\mu} &=& \frac{1}{2}\left( vV + wW -4C_s^2 + uU - 4C_d^2 - M^2 + L^2  +
Sym.\right) +8
\ena
where $Sym$ stands for the symmetrization of all the products appearing inside
the parenthesis.
These four quantities satisfy
\bea
{[} \hat{S},\hat{Q} {]} &=& -2i\ (u^2-v^2-w^2)\ \hat{R} \nonumber \\
{[} \hat{S},\hat{R} {]} &=& 2i\ \hat{Q} \nonumber \\
{[} \hat{Q},\hat{R} {]} &=& -8i(v^2+w^2)\ \left((u^2-v^2-w^2)\, \hat{\mu}
-2\hat{S}^2\right)\ \hat{S}
\label{eq:22}
\ena
We remark that, as conjectured in the general case, the r.h.s. in
(\ref{eq:22}) are polynomials.
When performing a "renormalization" of these four quantities:
\bea
{S} &=& \frac{1}{\sqrt{u^2-v^2-w^2}}\ \hat{S} \label{eq:21a} \\
{R} &=& \frac{1}{\sqrt{u^2-v^2-w^2}\,\sqrt{v^2+w^2}}\ \hat{R}
\label{eq:21b} \\
{Q} &=& \frac{1}{(u^2-v^2-w^2)\sqrt{v^2+w^2}}\ \hat{Q} \label{eq:21c} \\
\mu &=& \hat{\mu}
\label{eq:21d}
\ena
we recover the algebra obtained in \cite{5}:
\bea
{[} {S},{Q} {]} &=& -2i{R} \nonumber \\
{[} {S},{R} {]} &=& 2i{Q} \\
{[} {Q},{R} {]} &=& -8i{S} (\mu -2{S}^2) \nonumber
\ena
One recognizes also the "deformed" $g \ell(2)$ algebra, already considered in
\cite{3,4}.

Two remarks need to be added. First, we note that the generators of this
$g\ell(2)$-like algebra as given in (\ref{eq:20}) are not all obtained as
polynomials in the $sp(4)$ generators. As discussed in section \ref{s2.2}
one can,
in order to recover the polynomiality property of $W$ generators obtained via
Hamiltonian reduction in the highest weight gauge, consider the enlarged
\cws\ algebra by including the elements $J_u, J_v, J_w$ forming the
set $\cg^*_-$.
As already explained, the quotient $\cws/\cg^*_-$
leads to a realization
of the $\cw(sp(4), 2s\ell(2))$ algebra, whose
quantization (section \ref{s2.3}) provides the realization
(\ref{eq:21a}-\ref{eq:21d}).

Secondly one can note that any state in a $sp(4)$ representation is completely
specified by six numbers, each eigenvalue of one of six simultaneously
commuting operators: usually are considered the two $sp(4)$ fundamental
invariants $\Delta^{(2)}$ and $\Delta^{(4)}$, then one can choose the two
Casimir and two Cartan generators of the two $s\ell(2)$
forming the maximal $s\ell(2)
\oplus s\ell(2)$ subalgebra in $sp(4)$.
{}From the algebra
(\ref{eq:22}), $\mu, S$ and the $g\ell(2)$-like Casimir operator
\be
\Delta^{Q,R,S} = Q^2 + R^2 + 4(\mu-4) S^2 -4S^4
\label{eq:23}
\ee
can obviously be diagonalized simultaneously,
which, joined to the three generators $u,v,w$ constitute a complete set of
commuting observables as could be expected. Note also that the $sp(4)$
fundamental invariants can easily be recognized in this basis via the relations
\cite{5}:
\be
\Delta^{(2)} = \frac{1}{24} (\mu-8) \ \ ; \ \  \Delta^{(4)} - \frac{1}{6}
(\Delta^{(2)})^2 = \frac{1}{6912} (\Delta^{QRS} + 8\mu -64)
\label{eq:24}
\ee

Thus, finite \cw-algebras can occasionally be used for the determination in a
given simple Lie algebra of complete sets of commuting observables.

Now we are in position to generalize the $(x_i, p_i)$ representation of
$sp(4)$ given in (\ref{eq:10}-\ref{eq:12}) to a more general one with the
help of the $S,Q,R,\mu$ generators. As in the $sp(2)$ previous case, we wish
to have a coordinate representation. So we will keep unchanged $u,v,w$ as in
(\ref{eq:10}). In order to follow
closely the notations of \cite{5} and facilitate the discussion, let us also
introduce the parametrization:
\be
u=r^2 \ \ \  v=r^2 \sin \theta \cos2 \phi \ \ \  w=r^2 \sin \theta \sin2 \phi
\label{eq:25}
\ee
with $0 \leq \theta \leq \frac{\pi}{2}$, $0\leq \phi\leq\pi$, and also define:
\bea
C_\rho &=& \ \ \ \cos (2\phi) \ 2C_d + \sin (2\phi) \ M \nonumber \\
C_\phi &=& - \sin (2\phi) \ 2C_d + \cos (2\phi) \ M
\label{eq:26}
\ena
The simplest quantity to examine is
\be
\sqrt{u^2 -v^2 -w^2} \ S = -u L + \sqrt{v^2 + w^2} \ \ C_\phi .
\label{eq:27}
\ee

We immediately note that for $S=0$, one has the relation
\be
uL = \sqrt{v^2+w^2} \ C_\phi
\label{eq:28}
\ee
in accordance with the $(x_i, p_i)$ representation expressed in
(\ref{eq:10}-\ref{eq:12}).

Let us decide to keep $L$ unchanged. Then, for $S \neq 0, C_\phi$ becomes:
\be
C_\phi \rightarrow C'_\phi = \frac{uL}{\sqrt{v^2+w^2}} +
\frac{\sqrt{u^2-v^2-w^2}}{\sqrt{v^2+w^2}} \ S\ =\ -\frac{i}{\sin \theta}
\prt_\phi + \mbox{cotg} \theta\ S
\label{eq:29}
\ee

After some reordering of the $sp(4)$ generators via their CR's, we can express
the quantities $Q,R,\mu$, denoted $X_i\ (i=1,2,3)$ as follows:
\be
X_i = f_i (u,v,w) \ U + g_i(u,v,w) \ V + h_i(u,v,w)\
W + k_i (u,v,w,L,C_s,C_\rho,C_\phi)
\label{eq:30}
\ee

Using formula (\ref{eq:29}) for $C_\phi$, and keeping $L,C_s, C_\rho$ unchanged
(as well as of course $u,v,w$), one directly gets the general coordinate
realization of $U,V,W$, as given in \cite{5} (eq. (75-78)). We just rewrite
for illustration the $U$ generator expression:
\bea
U &=& \frac{1}{r^2} \left[ -\prt_r r^2 \prt_r -
\frac{4}{\sin \theta} \prt_\theta \sin \theta \prt_\theta +
(C_\phi)^2 + S^2 + \frac{\mu - 2 S^2 + \sin \theta Q}{\cos^2 \theta}
+ \frac{3}{4} \right]
\label{eq:31}
\ena

Thus the four operators associated with the finite \cw-algebra are
naturally used to
obtain, after modification of the expressions of four $sp(4)$ generators, the
general "coordinate" realization of $sp(4)$. As developed in \cite{5}, matrix
representations of the $g\ell(2)$-like algebra,
in which the wave functions behave
as vectors, have to be considered.

\indent

\subsection{Two particles in $d>2$}

\indent

The above construction can directly be generalized to dimensions $d>2$. There
the quadratic homogeneous polynomials in $x_i, p_i\ (i=1...d)$ generate the
symplectic
algebra $sp(2d,\R)$. One recognizes $\shalf d(d+1)$
coordinate generators $x_i x_j\ (i,j=1, ...d)$,
$\shalf d(d+1)$ ones relative to the $p_ip_j$ second order
differential operators, and $d^2$ ones associated to the $x_ip_j$. Again one
remarks the gradation:
\be
{\cg} = {\cg}_{-1} \oplus {\cg}_0 \oplus {\cg}_1
\label{eq:32}
\ee
with the Abelian subalgebras ${\cg}_{\pm 1}$ generated by the $x_i x_j$ and
$p_ip_j$ quantities respectively, while ${\cg}_0$, isomorphic to $g\ell(d,\R)$,
is spanned by the $x_ip_j$'s. Such a gradation is relative to
the diagonal $s\ell(2)$ in the regular
subalgebra direct sum of $d\ s\ell(2)$ ones.
The associated finite \cw-algebra, will
be of dimension $d^2$ and a "deformed" version of the $g\ell(d)$
algebra in the same way we have found in $sp(4)$ an $g\ell(2)$-like algebra
(\ref{eq:22}). More precisely, from the $s\ell(2)$
decomposition
\be
\cg= \frac{d(d+1)}{2} D_1 \oplus\frac{d(d-1)}{2} D_0
\ee
it is clear that the \cw-algebra will contain a Lie
subalgebra of dimension $\frac{d(d-1)}{2}$, and that
the other $\frac{d(d+1)}{2}$ $W$ generators will close with
cubic (or lower degree) terms. When going from the bosonic representation to a
general one, we will express the $sp(2d)$ generators as functions of the
$x_i$'s, $p_i$'s, and $W$ generators. The
set of observables will be constituted by the $\frac{d(d+1)}{2}$
$x_ix_j$'s and the $d-1$ Casimir operators of the "deformed $g\ell(d)$"
algebra, together with its $d$ Cartan generators. We recover the
right number of operators in order to form a complete set
of commuting observables for
$sp(2d)$.
Note that
the second order differential operators ($\cg_1$ part of
$sp(2d)$) exhaust the number of $W$ generators only when $d=1$, while
in other cases, we will have to modify also $\frac{d(d-1)}{2}$
zero grade generators.

\subsection{Anyons in $d=2$ ?}

\indent

It is reasonable to expect the $d=2$ case to be a generalization of the $d=1$
one, in which the anyonic parameter is clearly identified. However, the author
of \cite{5} had a rather  negative conclusion based on the following argument.
They found the boson and fermion assigned to representations of dimension
$J=K+1=1$ and 2 respectively of the \cw-algebra (\ref{eq:22}).
The parameter $K$ is
obviously discrete, which ruins an anyonic interpolation between bosons and
fermions.

\indent

We would like to discuss the following alternative.
First, we recall that the authors of \cite{5} works with multivalued
functions. We choose to work with univalued ones. Then,
the operator $L$ introduced in the previous section
\be
L=-i\frac{\prt}{\prt\phi}  \label{Lo}
\ee
is not well-defined on the set $L_2([0,\pi])$. More precisely, for $L$
to be self-adjoint, we have to:

- either restrict it on functions satisfying
\be
\psi(0)=\lambda\psi(\pi)
\ee
where $\lambda$ plays the r\^{o}le of an anyonic parameter ($\lambda=1$
characterizes the bosons, while $\lambda=-1$ characterizes fermions)

- or modify the explicit form of $L$
\be
L= -i\frac{\prt}{\prt\phi} +\alpha \label{L1}
\ee
and apply it on functions such that
\be
\psi(0)=\psi(\pi) \label{bos}
\ee
the anyonic parameters being here $\alpha$ ($\alpha=0$ being the
bosons, and $\alpha=1$ being the fermions) \cite{Min}.

As we start with the bosonic realization, it is natural to assume that
we are working with functions of kind (\ref{bos}). Then, the form
(\ref{Lo}), compared with (\ref{L1}) "confirms" that we are really
working with bosons.
However, this is true only for the $x_i,p_j$ representation. When we
study a more general representation, we have to choose which
generators will be modified (see section \ref{sN2d2}). Until now,
following \cite{5}, we had chosen to keep $L$ unchanged.
We present hereafter some scenarii that may help to recognize anyons
in this framework.

\indent

$(i)$ Let us
choose to keep $C_\phi$ unchanged, and transform $L$. Then, from
(\ref{eq:27}), we get a new expression:
\be
L= -i\frac{\prt}{\prt\phi} -S\cos\theta \label{Lany}
\ee
By a computation analogous to the one performed in section \ref{sN2d2}, one
can realize the other generators of $sp(4,\R)$ in accordance with
(\ref{Lany}).
If we concentrate on the one-dimensional representations of the
\cw-algebra, $S$ becomes a number which can be non-zero.
Actually, it has been
shown in \cite{5} that the positivity of the two operators
$A_i=(p_i)^2+(x_i)^2$ $(i=1,2)$ imposes $S=\pm2$.
Then, the real number
$\alpha=S\cos\theta$ could be interpreted as the anyonic parameter
($\alpha=0$ corresponding to the usual boson and fermion cases).
Note however that the term $\cos\theta$ is not constant on any
$s\ell(2)$
subalgebras built from $L$, since it does not commute with the other
$s\ell(2)$ generators. This means that we have to break the
$s\ell(2)$ symmetry to recognize anyons.
For
higher dimensional representations, $S$ is a diagonal matrix, and we
have a generalization of anyons directly related to the enlarged
\cw-algebra (we recall that $\cos\theta=\sqrt{u^2-v^2-w^2}\,/ u$).

\indent

$(ii)$ One may think of not breaking the whole $sp(4)$ symmetry. Then,
one can choose for $L$:
\be
L= -i\frac{\prt}{\prt\phi} -(r^2\cos\theta)^2 S
\ee
The term $(r^2\cos\theta)^2$ commutes with the full $sl(2)$ algebra
generated by $L,M,2C_d$. Indeed, setting
\be
\rho_1=r^2\cos\theta \mb{and} \rho_2=r^2\sin\theta
\ee
one can rewrite $M$ and $2C_d$ in the bosonic representation as
\bea
M &=& -2i\frac{r^2}{\rho_2}\,
\left(\cos(2\phi)\, \rho_2\frac{\prt}{\prt\rho_2}
+\sin(2\phi)\, \frac{\prt}{\prt(2\phi)}\right) \\
{2C_d} &=& -2i\frac{r^2}{\rho_2}\,
\left(\sin(2\phi)\, \rho_2\frac{\prt}{\prt\rho_2}
-\cos(2\phi)\, \frac{\prt}{\prt(2\phi)}\right) \\
&\mbox{with}& r^2=\sqrt{\rho_1^2+\rho_2^2}
\ena
which clearly shows that $r^2\cos\theta\equiv\rho_1$ commutes with the
whole $s\ell(2,\R)$ subalgebra, and therefore is constant on each
$s\ell(2)$ representation.

In a finite dimensional $s\ell(2)$ representation, $L$ has only
(half-)integer eigenvalues.
In order to get a continuous parameter, we have to consider
infinite dimensional $s\ell(2)$-representations. Then
$\rho_1$ could be seen as a continuous free parameter, and
$\rho_1 S$ interpreted as a generalized anyonic parameter.
In this context, focussing on the one-dimensional representation of
the \cw-algebra,
a given $s\ell(2)$ representation would define one anyon type. In
the complete $sp(4)$ representation we recognize all the anyons, two
anyonic $s\ell(2)$ representations being related by a $sp(4)$
transformation. Such a situation generalizes to higher dimensional
\cw-representations.

\indent

$(iii)$ Finally, one chooses
\be
L=-i\frac{\prt}{\prt\phi} -S
\ee
We still have to take infinite
dimensional representations for the $sl(2)$ subalgebra. Considering the
subalgebra generated by $L, V+v, W+w$, it is enough to release the
positivity of the spectrum for $A_1-A_2$ (i.e. ask the spectrum of
$A_1+A_2$ only to be positive instead of the spectrum of both $A_1$ and
$A_2$) to allow infinite dimensional representations.

\sect{Conclusion}

\indent

We have presented a construction of finite \cw-algebras in which $W$
elements are functions of all the current components $J_a$, in
opposition to the usual case where some of these quantities are
assigned to constants. Such an approach has been developed when the
associated symmetry subgroup $H$, w.r.t. which are determined the $J$
invariant quantites, is Abelian. Our construction can be generalized
for a more general $H$, or in other words for any $s\ell(2)$
embeddings in \cg. The extension to the affine case can also be
considered. These works are in progress and will be presented
elsewhere \cite{prep}.

Then, we have recognized a class of such finite \cw-algebras in the
framework of Heisenberg quantization for $N=2$ identical particles.
{}From the explicit expressions of the $W$ generators, one can deduce in
$d$ dimensions a general realization of the observable algebra under
consideration. Moreover, the anyonic parameter being directly related
to the unique $W$ generator in the $d=1$ case, we have tried to
identify it again in the larger \cw-algebra relative to the $d=2$ case.
Interpretation of the whole algebra still needs to be precised. Of
course, the case of $N>2$ identical particles, in which the algebra of
observables becomes infinite \cite{5} deserves a careful study and
retains now our attention.

\indent

{\Large {\bf Acknowledgements}}

\indent

We are endebted to H. H\"{o}gaasen for prompting our attention to the
reference \cite{5}, which plays a crucial r\^{o}le in this work, and for
encouragements.
It is a pleasure to thank F. Delduc, L. Frappat and M. Mintchev for
important comments, and L. Feher, G. Girardi,
J.C. Le Guillou, and A. Liguori for stimulating discussions.

\indent


\begin{thebibliography}{99}
\bibitem{1} T. Tjin, \PL{B292} (1992) 60.
\bibitem{2} J. de Boer and T. Tjin, \CMP{158} (1993) 485.
\bibitem{3} P. Bowcock, Preprint DTP 94-5.
\bibitem{4} P.W. Higgs, Journ. Phys. A, Math. Gene. {\bf 12} (1979) 309.
\bibitem{5} J.M. Leinaas and J. Myrheim, \IJMP{A8} (1993) 3649.
\bibitem{Bais} F.A. Bais, T. Tjin, P. van Driel, \NP{B357} (1991) 632.
\bibitem{ORaf} L. Feher, L. O'Raifeartaigh, P. Ruelle, I. Tsutsui and
A. Wipf, Phys. Rep. {\bf 222} (1992) 1, and references therein.
\bibitem{U1} F. Delduc, E. Ragoucy, P. Sorba, \PL{B279} (1992) 319.
\bibitem{FRS} L. Frappat, E. Ragoucy, P. Sorba, \CMP{157} (1993) 499.
\bibitem{thTjin} T. Tjin, {\it Finite and infinite W algebras and
their applications}, PhD thesis.
\bibitem{Ler} For a recent review, see A. Lerda {\it Anyons}, Lecture Note in
Physics, Monographs {\bf m 14}, Springer-Verlag 1992.
\bibitem{Min} M. Mintchev, private communication.
\bibitem{prep} F. Barbarin, F. Delduc, E. Ragoucy and P. Sorba,
in preparation
\end{thebibliography}
\end{document}